\documentclass[cameraready]{Interspeech}
\usepackage{siunitx}
\usepackage{tikz}
\usepackage{everypage}
\newcommand\BackgroundText{%
  \begin{tikzpicture}[remember picture,overlay]
    \node [rotate=0, scale=1.3, text opacity=1.0, color=gray] at (current page.south west) [anchor=south west, xshift=0.7cm, yshift=1.9cm] {accepted at Interspeech 2026};
  \end{tikzpicture}
}
\AddEverypageHook{\BackgroundText}

\title{Not Quite My Tempo: \\Voice Activity-aware Speech Synthesis for Lip-Synchronous Dubbing}

\author[affiliation={1}, orcid=0000-0002-0453-2114, correspondingauthor]{Alejandro}{Pérez-González-de-Martos}
\author[affiliation={1}, orcid=0000-0001-9712-6037]{Florian}{Lux}
\author[affiliation={1}]{Angelina}{Elizarova}
\author[affiliation={1}]{\\Milana}{Shkhanukova}
\author[affiliation={1}]{Andreas}{Kellner}
\author[affiliation={1}, orcid=0000-0003-4491-0025]{Mattia Antonino}{Di Gangi}

\address{
    $^1$ AppTek GmbH, Germany 
}

\email{aperez@apptek.com, flux@apptek.com}


\begin{document}

\maketitle

\begin{abstract}
    Automatic lip-synchronous dubbing requires a speech synthesis model to generate alternating voice and silence patterns in the target language that match the timing of the source clip precisely to ensure an optimal viewing experience. Prior works address this problem by conditioning the speech synthesis process on lip movements extracted from the video signal. In this work, we condition the speech generation on a binary voice-activity signal, which has a lightweight representation and can be produced in multiple ways. We show that the model follows the voice-activity signal with high accuracy while maintaining natural prosody and semantically appropriate pause placement within sentences, as demonstrated through extensive objective and subjective evaluations. By randomly masking this condition during training, we make the feature entirely optional during inference, allowing editors to enforce or relax lip-sync constraints when desired. 
\end{abstract}

\section{Introduction}

While modern text-to-speech (TTS) synthesis has evolved through paradigms like regression, next-token prediction, transport functions or inpainting, the application of these methods to automatic dubbing introduces unique challenges.
The goal extends beyond simple translation to the creation of a seamless cross-lingual experience. Consequently, alignment must be multidimensional, encompassing not only voice identity, style and intent, but interestingly and uniquely, also the precise temporal mapping of pause structures. The temporal axis of the target speech must be precisely mapped to the source to maintain the visual-audio coherence required for lip-sync dubbing.



To this end, we propose an approach that incorporates a Voice Activity Detection (VAD) stage to generate frame-level binary masks. These voice activity annotations provide the temporal conditioning necessary to condition an inpainting-based synthesis process~\cite{chen2025f5, zhu2025zipvoice} during training. At inference time, the model continues to leverage the VAD-derived mask, which is now derived from the reference audio in the source language. Owing to the inpainting paradigm operating on a fixed-length canvas, the binary voiced/unvoiced pattern can be seamlessly propagated from the source signal to the target, since both are defined over the same number of frames. This design constrains only the temporal structure of speech activity, while leaving the allocation of linguistic content within voiced regions to the model, which is trained end-to-end. Notably, although no explicit constraints on pause placement are imposed beyond the activity mask, we find the model consistently learns to position pauses at linguistically coherent boundaries. To illustrate this, we provide example audios taken from our subjective evaluation, as well as examples from our qualitative analysis.\footnote{\url{https://anondemos.github.io/NotQuiteMyTempo/}} 


\begin{figure}[t]  
    \centering
    \includegraphics[trim={0 13cm 9cm 0},clip,width=\columnwidth]{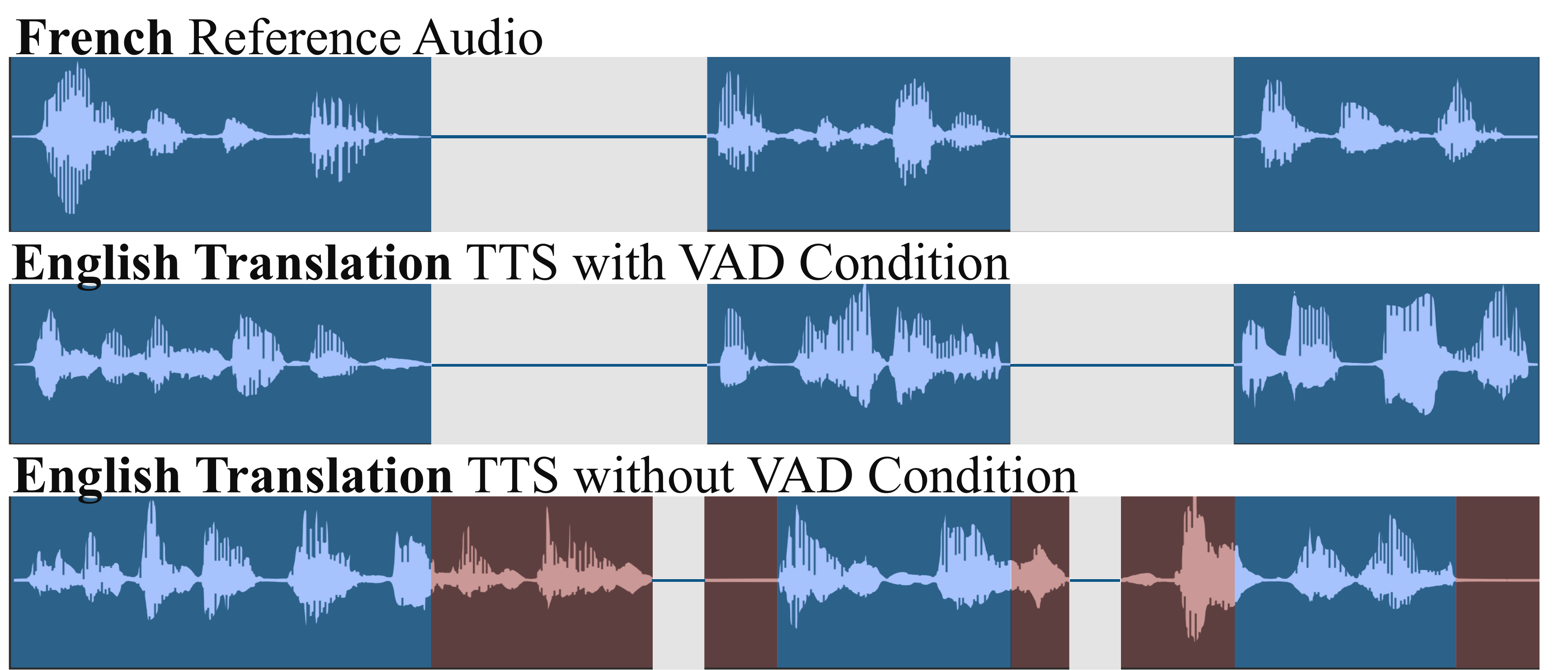}
    \caption{Example of our proposed technique in action: The VAD condition encourages the model to use only the blue area as canvas for inpainting-TTS. Silences are aligned on the time axis, which is a requirement in lip-synchronous dubbing.}
    \label{fig:teaser}
\end{figure}

While numerous works have explored methods for transferring voice timbres and speaking style \cite{jia2018transfer,wang2018style,ZhangWZWCSJRR19,wu2022adaspeech}, few investigated the problem of achieving precise temporal alignment between source and target utterances. Exact prosody matching is explored in research fields such as voice privacy \cite{meyer2023prosody}, deepfake detection \cite{wang2026asvspoof, lux2023exact}, and literary studies \cite{koch2022poetictts}, but these settings assume identical source and target texts. This assumption does not reflect the challenges of dubbing.

Most existing approaches for prosodic alignment across differing utterances in dubbing and similar contexts attempt to address this challenge by leveraging visual cues from a video stream, for example by encoding lip movements into embeddings that condition the text-to-speech system \cite{hu2021neural,sahipjohn2024dubwise,wang2025syncvoice}. In contrast, our method relies solely on audio-text data and does not require paired video during training, eliminating a major practical constraint. Additionally, our approach is decoupled from the problem of detecting the primary speaking face and modeling its lip dynamics, a process that can be brittle in out-of-domain scenarios such as cartoons, anime, or scenes with multiple visible speakers, and may require specialized encoders.

Effective prosody transfer in lip-sync dubbing requires translations that accommodate the pacing and length of the original speech. In traditional professional workflows, human experts manually adapt the translated script to satisfy these rigid lip-sync constraints, but recent research has sought to automate the process through dubbing-specific machine translation (MT). Isochronous MT aims at generating target text that matches the source duration by aligning syllable or phoneme counts \cite{lakew-etal-2019-controlling, tam22_interspeech, wilken-matusov-2022-appteks}. Additionally, some works focused on transferring explicit text pause markers from source to target to provide a richer signal to speech synthesis~\cite{federico-prosodic-alignment}. These lines of work are complementary to ours, as a speech synthesis system still needs to render the written translation according to the dubbing requirements. In the present work, we assume that a given translation has a sufficiently matching timing structure, but we do not require explicit additional prosodic information.

\section{Methods}

\subsection{Basic TTS Setup} 
\begin{figure}[b]
  \centering
  \includegraphics[trim={0 0 9cm 4cm},clip,width=\columnwidth]{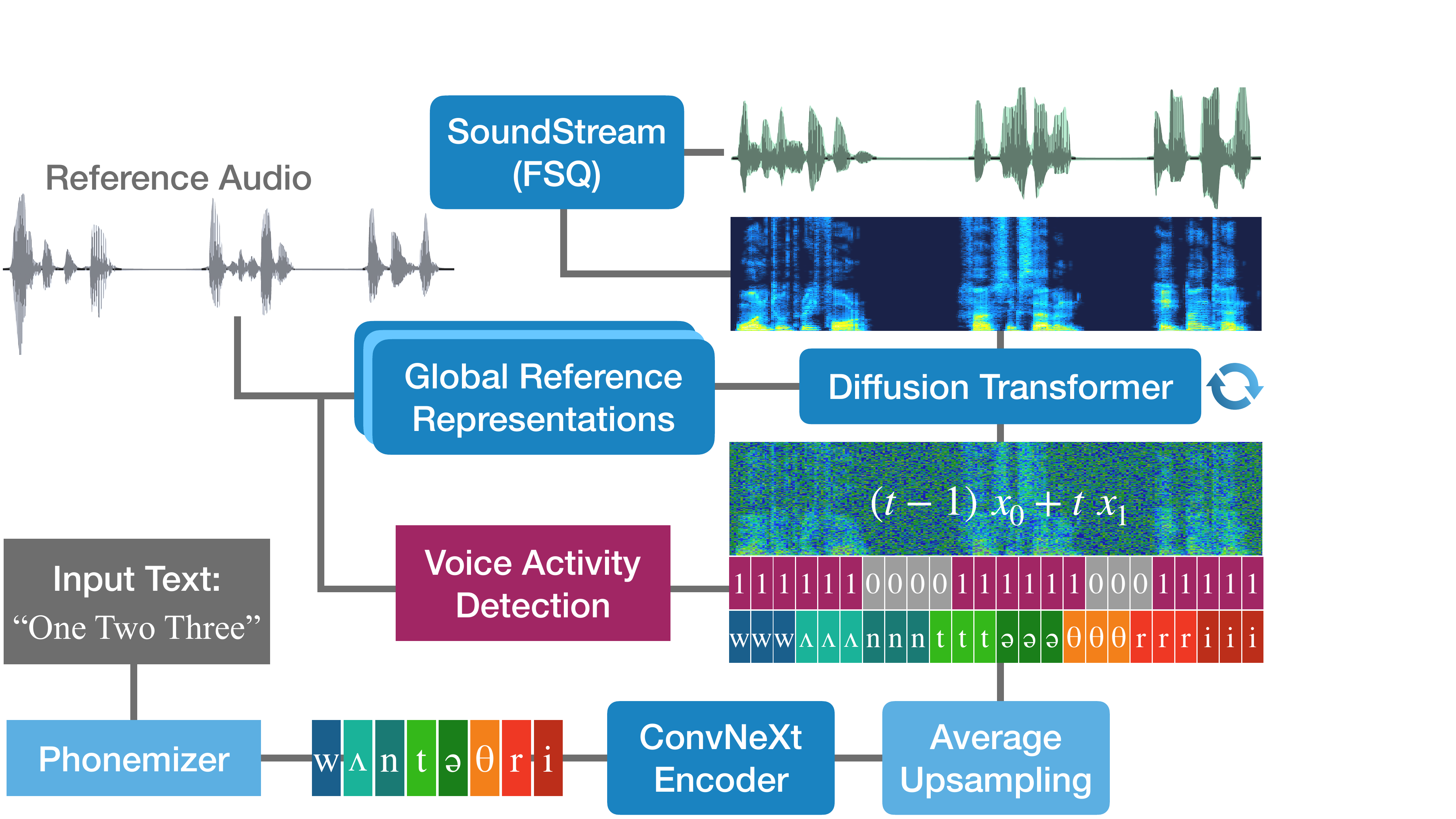}
  \caption{Overview of the proposed architecture with VAD conditioning. During training, VAD labels are randomly masked, enabling optional use of this feature at inference time.}
  \label{fig:sys_overview}
\end{figure}

Our model architecture loosely follows F5-TTS \cite{chen2025f5, eskimez2024e2} with a few modifications. Figure~\ref{fig:sys_overview} provides a schematic overview of the system. First, we replace filler-token-based upsampling with the average upsampling method proposed in ZipVoice \cite{zhu2025zipvoice}, which provides a stronger inductive bias toward near-diagonal temporal alignment between input and target sequences. Second, our model addresses zero-shot speaker and style transfer via explicit speaker embeddings \cite{jia2018transfer} and Global Style Tokens (GST) \cite{wang2018style} as opposed to acoustic prompt-based conditioning (inpainting), leading to improved disentanglement of timbre, style, and accent. Finally, we adopt a pretrained, modified SoundStream \cite{zeghidour2021soundstream} vocoder which maps 16 kHz waveforms to 32-dimensional scalar-quantized latent codes \cite{yang2025simplespeech} and generates 48 kHz high-fidelity outputs \cite{liudelightfultts}.

\subsection{Voice Activity Conditioning} \label{section:vac}
The main novelty of this method is the introduction of a voice activity mask to condition audio generation. For each acoustic frame, voice activity is encoded as a binary indicator of speech presence or absence. This signal is embedded and added to the encoder representations, providing the Diffusion Transformer (DiT) \cite{peebles2023scalable} Flow Matching decoder with explicit temporal voice activity cues. A schematic illustration is provided in Figure~\ref{fig:sys_overview}.

To enable controllability of voice activity adherence at inference time, we randomly mask a subset of the ground-truth voice activity embeddings during training. This allows selective masking of frames around speech–silence transition regions, which can be useful in post-editing workflows. Fully masking the voice activity signal effectively removes VAD conditioning from the model, allowing the system to operate both with and without VAD guidance. This flexibility is motivated by the observation that strict lip-syncing constraints are not required in all scenarios: one large-scale study reports that such alignment is present in movies only about 12\% of the time~\cite{brannon2023dubbing}. Moreover, accommodating non-adapted translations (e.g., standard MT outputs) under rigid lip-sync constraints is inherently difficult. Our results further indicate that, in some cases, enforcing such alignment may compromise other aspects of voice quality.

\section{Experiments}

\subsection{Data} \label{sec:data}
We provide results for two models: a core English model trained on the publicly available LibriTTS-R corpus \cite{zen2019libritts, Koizumi2023} to ensure reproducibility, alongside a multilingual model trained on a larger combination of public and proprietary data sources.

For evaluation, we utilize a subset of the Multilingual TEDx (mTEDx) dataset \cite{salesky2021multilingual}. This corpus provides diverse source-to-English language pairs for comprehensive cross-lingual assessment. Moreover, the semi-spontaneous nature of TEDx talks, characterized by frequent and irregular pauses, offers a more rigorous test than traditional read-speech corpora. Leveraging such challenging dynamic speech patterns allows us to effectively validate our approach's precision in managing timing constraints under realistic conditions. 
To this end, we curated a test set of utterances with durations between 7 and 15 seconds, each containing at least one pause exceeding 500 ms. The final evaluation set comprises 291 samples, balanced across four source languages: Greek, French, Portuguese and Russian. For subjective assessments, a random subset of 25 samples was selected from this pool; this sample size was constrained to maintain a reasonable evaluation time and mitigate rater fatigue.

\subsection{Experimental Setup}
The proposed model operates on phoneme sequences, which are mapped to 1024-dimensional embeddings and contextualized via an 8-layer ConvNeXt encoder. The encoder outputs are upsampled to the target sequence length using the average upsampling method from ZipVoice. 
We perform cross-lingual voice and style transfer using explicit conditioning embeddings. To condition the model on the voice-timbre of a reference we employ a stack of pretrained speaker encoders (FACodec~\cite{ju2024naturalspeech} and ERes2NetV2~\cite{chen2024eres2netv2}), while prosodic and emotional cues are captured by a GST-based encoder~\cite{wang2018style}. These embeddings are added to the DiT blocks using Adaptive Layer Normalization~\cite{peebles2023scalable}. To ensure robustness against noisy references, we apply data augmentation to the GST encoder inputs during training.


Frame-level voice activity information is extracted using the Silero VAD model.\footnote{\url{https://github.com/snakers4/silero-vad}} As discussed in Section~\ref{section:vac}, we apply random masking to the voice activity embeddings during training to enhance model controllability. The vector field is parametrized by an 18-layer DiT decoder ($d = 1024$, 8 attention heads) and optimized via the optimal-transport Conditional Flow Matching objective~\cite{lipmanflow, mehta2024matcha}.
All models are trained using a global batch size of 128 across four NVIDIA A100 GPUs. We use a peak learning rate of \num{7e-5} with a linear warmup of 10k steps, decayed to \num{5e-6} over 800k steps. 
To enable classifier-free guidance (CFG) at inference time \cite{ho2021classifier}, conditioning inputs are dropped with a probability of 20\% during training.

\subsection{Alignment Consistency}
To assess the model's ability to adhere to voice activity masks, we compute frame-level VAD accuracy metrics between the reference and synthesized audios from the mTEDx test set. Table~\ref{tab:vad_lang} presents frame-level VAD alignment accuracy scores for various source languages under both conditioned and unconditioned settings. Across all languages, regardless of each language's unique features, VAD conditioning significantly improves alignment performance, increasing accuracy from roughly 73\% without conditioning (overlap by chance) to about 96\% with conditioning for the model trained on LibriTTS-R, and similarly for our multilingual model. These findings confirm that explicit VAD guidance is highly effective in enforcing temporal alignment. We hypothesize that the slightly lower accuracy of the multilingual model stems from the increased difficulty of VAD in our multilingual dataset due to frequent non-speech events occurring, such as e.g. hesitations or laughter, as well as other voice modes, such as e.g. yelling and whispering.

\begin{figure}[t]
  \centering
  \includegraphics[width=\columnwidth]{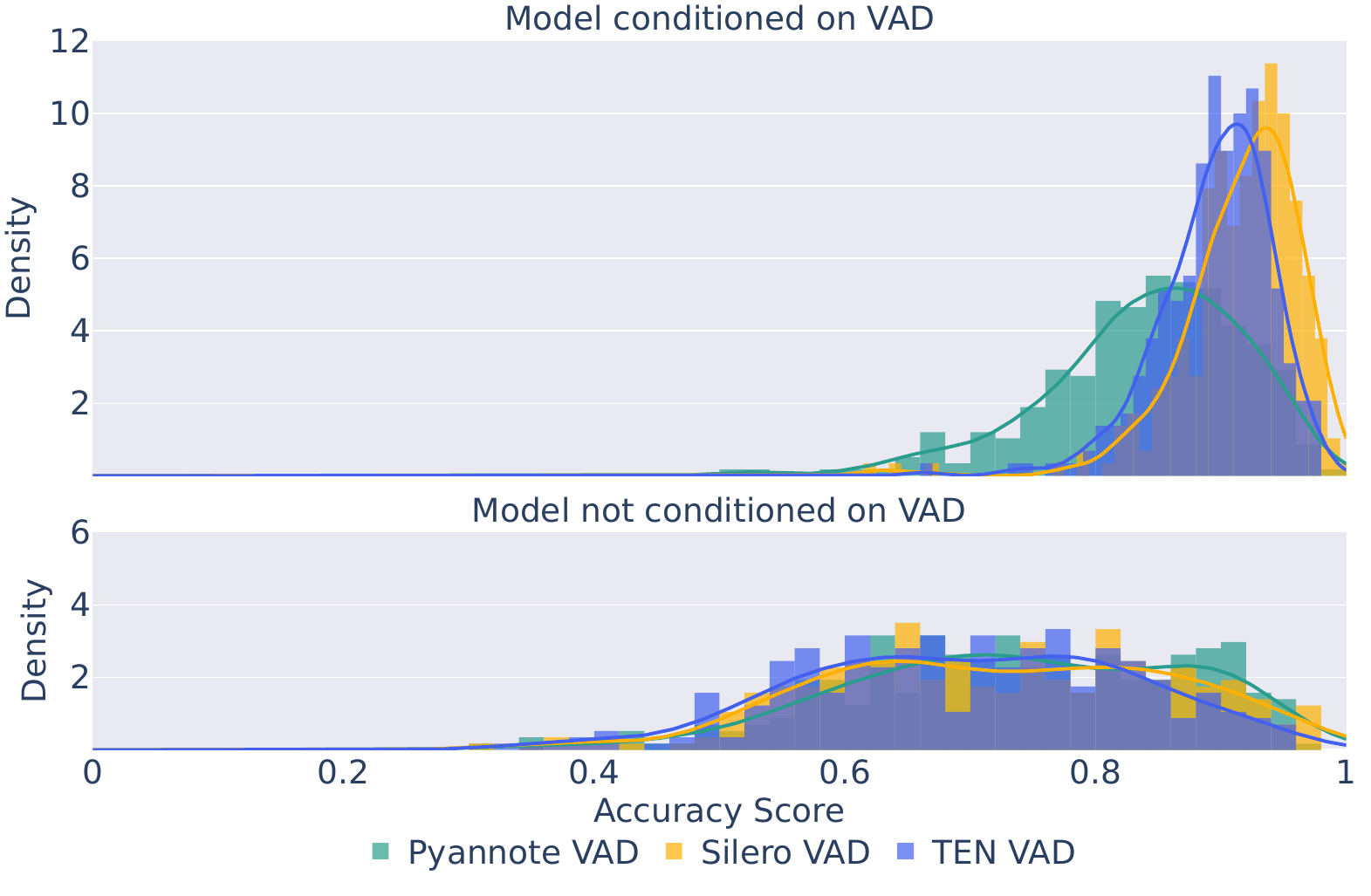}
  \caption{Comparison of the distribution density of the accuracies per sample, measured by different VAD models.}
  \label{fig:acc_density}
\end{figure}

To ensure the accuracy evaluation is not overly influenced by a particular VAD system, we use two additional state of the art VAD models to measure the accuracy of the model outputs. Figure~\ref{fig:acc_density} shows the density distributions of the accuracy scores for both conditioned and unconditioned synthesis across Pyannote~\cite{bredin2020pyannote}, TEN\footnote{\url{https://github.com/TEN-framework/ten-vad}}, and Silero. While these can vary in sensitivity, the conditioned synthesis (top) demonstrates consistently higher accuracy scores across all three VAD systems.

\begin{table}[b]
\centering
\caption{Silero VAD alignment accuracy per source language, with and without VAD conditioning. No VAD accuracy shows overlap by chance in each language.}
\label{tab:vad_lang}
\begin{tabular}{lcccc}
\toprule
& \multicolumn{2}{c}{LibriTTS Model} & \multicolumn{2}{c}{Multilingual Model}  \\
\midrule
\textbf{Source} & \textbf{No VAD} & \textbf{VAD} & \textbf{No VAD} & \textbf{VAD} \\
\midrule
Greek & 66.95\%  & 96.06\% & 66.06\%  & 91.28\%\\
French & 68.65\%  & 96.09\% & 68.80\% & 92.21\% \\
Portuguese & 75.17\%  & 95.97\% & 73.06\% & 91.67\% \\
Russian & 81.17\% & 96.74\% & 78.50\% & 91.14\% \\
\midrule
\textbf{Total} & 72.69\% & 96.21\%  & 71.35\% & 91.59\% \\
\bottomrule
\end{tabular}
\end{table}

\begin{figure}[t]
   \centering
   \includegraphics[trim={1.0cm 0cm 1.5cm 0cm},clip,width=\columnwidth]{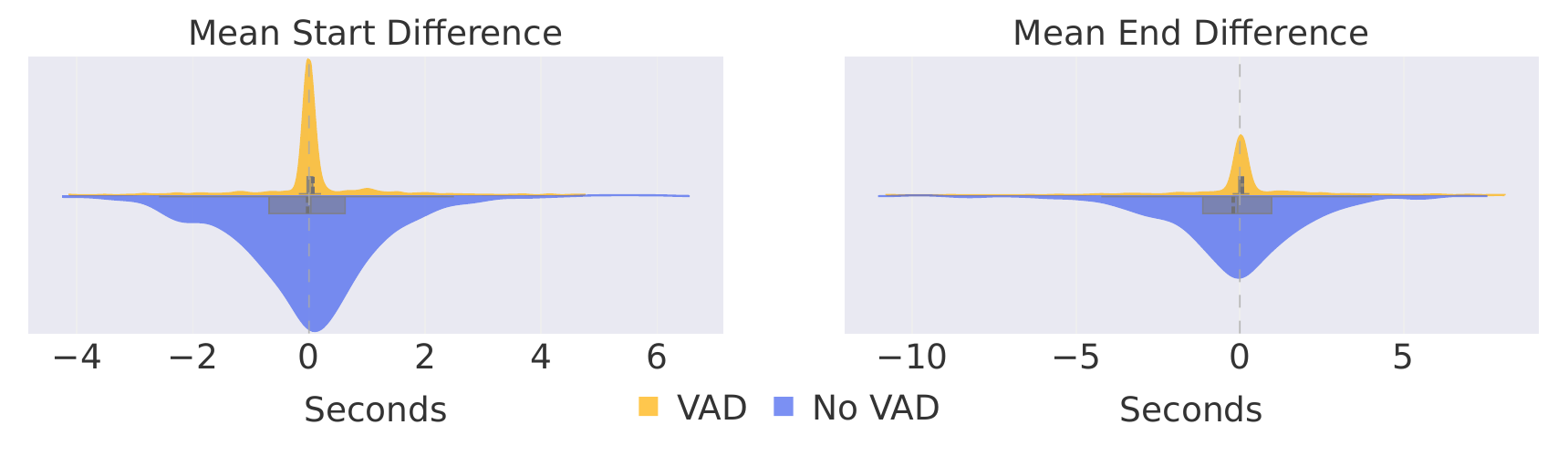}
   \caption{Visualizing the deviation on the time axis of silence starts and silence ends, averaged across each sample.}
   \label{fig:distribution}
 \end{figure}

Precise onset and offset timings are critical for high-quality dubbing. 
%
We evaluate this by measuring the temporal offsets of silence boundaries between the source and synthesized speech.
Figure~\ref{fig:distribution} shows the distribution of these timing offsets, averaged across the evaluation set.
When conditioned on VAD, the boundary deviations are tightly concentrated around zero, indicating high temporal precision. In contrast, the model without VAD conditioning exhibits a wider spread across both positive and negative deviations. 

\subsection{Prosody Naturalness} \label{sec:subjective}
%
While VAD conditioning ensures temporal alignment, the model must also maintain semantic and prosodic coherence through appropriate pause placement and dynamic pacing to fit the target duration.
To evaluate these aspects, we performed a subjective evaluation with 40 English native speakers recruited via the Prolific crowd-working platform.\footnote{\url{https://www.prolific.com/}} Participants were compensated fairly according to high ethical standards. Using the subjective test set described in Section~\ref{sec:data}, we synthesized the English translations with and without VAD conditioning for both our LibriTTS and multilingual variants. 
%
Each participant evaluated 48 audio samples, derived from 12 randomly selected references (out of a pool of 25) across all four model configurations. Raters provided scores on a 5-point Likert scale regarding the naturalness of pause locations (Placement MOS) and the overall naturalness of pacing and intonation (Prosody MOS).

\begin{figure*}[t]
    \centering
    \includegraphics[width=\textwidth]{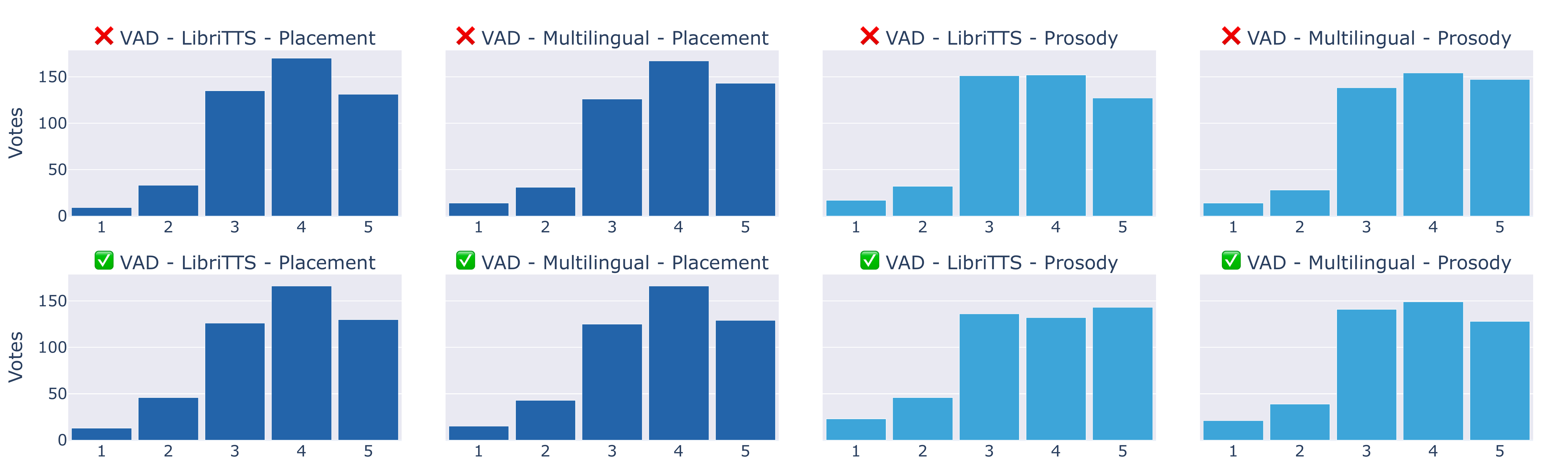}
    \caption{Subjective evaluation results. There are 3840 individual ratings from 40 participants, split uniformly across all systems and conditions. Placement MOS (darker, left) and Prosody MOS (lighter, right) denote naturalness of pause locations and overall prosody, respectively. Rows contrast baseline (top) with VAD-conditioned (bottom) configurations.}
    \label{fig:subjective_results}
\end{figure*}

The distribution of subjective ratings is illustrated in Figure~\ref{fig:subjective_results}, with aggregated MOS reported in Table~\ref{tab:subjective_results}. 
While a marginal decrease in both scores was observed when the VAD conditioning is enabled, the difference is not statistically significant according to pairwise Mann-Whitney~U tests \cite{mann1947test} ($p>0.05$ for all pairwise $p$). These findings suggest that VAD conditioning does not degrade the perceived naturalness of the prosody despite the added temporal constraints. Furthermore, it indicates that the model maintains the ability of placing pauses at semantically appropriate locations. We further investigate the outlier samples with the lowest rating scores in Section~\ref{sec:analysis}.

\begin{table}[b]  
        \caption{Mean Opinion Scores rated on a 5-point Likert scale with standard deviation. Each score is based on 40 ratings.}
    \centering
    \begin{tabular}{lccc}
        \toprule
        Model & Placement & Prosody  \\
        \midrule
        LibriTTS No VAD       & 3.80 $\pm$ 0.98 & 3.71 $\pm$ 1.04 \\
        LibriTTS VAD          & 3.74 $\pm$ 1.04 & 3.68 $\pm$ 1.14 \\
        \midrule
        Multilingual No VAD   & 3.82 $\pm$ 1.02 & 3.81 $\pm$ 1.03 \\
        Multilingual VAD      & 3.73 $\pm$ 1.05 & 3.68 $\pm$ 1.09 \\
        \bottomrule
    \end{tabular}
    \label{tab:subjective_results}
\end{table}

\subsection{Robustness} \label{sec:robustness}
%
The impact of VAD conditioning on synthesis robustness is assessed via Word Error Rate (WER) using an internal proprietary ASR system, complemented by Intelligibility and Prosody scores from the TTSDS benchmark suite~\cite{minixhofer2024ttsds, minixhofer2026ttsds}.
These results are summarized in Table~\ref{tab:objective_results}.
We observe a slight increase in WER for VAD-conditioned synthesis, which is primarily attributed to instances where the translated text is poorly adapted to the source VAD constraints. In such cases of extreme temporal mismatch, the model may aggressively adjust pacing or introduce phonetic deletions and repetitions to satisfy the enforced timing boundaries. A more detailed qualitative analysis of such edge cases and their associated error modes is provided in Section~\ref{sec:analysis}.

\label{sec:objective}
\begin{table}[b]  
        \caption{Synthesis robustness objective results: ASR-based WER and TTSDS benchmark Intelligibility and Prosody scores.} 
    \centering
    \begin{tabular}{lccc}
        \toprule
        Model & WER & Intelligibility & Prosody \\
        \midrule
        LibriTTS No VAD       & 8.5\% & 76.82 & 85.00 \\
        LibriTTS VAD          & 13.9\% & 80.59 & 86.98 \\
        \midrule
        Multilingual No VAD   & 6.3\% & 78.90 & 82.96 \\
        Multilingual VAD      & 8.3\% & 79.72 & 86.32 \\
        \bottomrule
    \end{tabular}
    \label{tab:objective_results}
\end{table}

\section{Qualitative Analysis} \label{sec:analysis}
This section investigates the performance trade-offs observed in Sections~\ref{sec:subjective} and~\ref{sec:robustness}. 
%
%
Looking at the worst-scoring samples, we find a strong correlation between low evaluation scores and the presence of synthesis artifacts, such as phonetic deletions, repetitions, or word reordering. Further analysis of samples with those defects reveals unique triggers for each error type.

%
Word omissions typically occur when suboptimal translations significantly exceed the available temporal budget.
%
Notably, the model prioritizes the integrity of the pause structure over verbatim synthesis.
When using constructed examples in which we gradually increase the number of words with a fixed-duration reference, the model 
can maintain alignment even at speaking rates exceeding the training distribution.
However, once a critical threshold is reached, the model preserves temporal synchronization by omitting parts of the sentence, ensuring that the alignment on the time axis remains unaffected.

%
Similarly, word repetitions are linked to translations with insufficient text content for the target duration.
%
Again, we observed that the model can reduce the speaking rate to maintain alignment without altering the VAD structure until a lower bound is reached.
Beyond this point,
the model 
eventually resorts to repetitions to occupy the remaining duration rather than further slowing down or inserting misaligned pauses.

%
In cases involving word reordering, we identified the presence of punctuation marks as a significant factor.
Further constructed examples confirmed that the model has learned a strong correlation between pauses and punctuation marks, such as commas and periods. Hence, whenever punctuation occurs near a VAD silence region, the model attempts to align the two by speeding up or slowing down the adjacent segments as needed. Similar to the previous cases, this only worked to some extent. If the speaking rate modulation became too extreme, 
the model preserves the VAD boundary by reordering the text rather than shifting the pause.

%
While unadapted translations may induce stability issues, these were successfully mitigated by manually refining the text to align with the source VAD structure. This confirms that the model's precision remains high when provided with isochronous input, either from automated isochronous MT or via human-in-the-loop intervention, to ensure that the translations are properly adapted to target pause structures.

\section{Conclusion}
We propose a modification to inpainting-based TTS, which allows for the automated creation of lip-synchronous dubs through the use of VAD as condition signal. Our experimental results show that the model follows the temporal structure of a reference audio with high accuracy. In subjective and objective evaluation of prosody and robustness, we observe only a minor degradation when using the proposed method, which we link to poorly adapted translations, unlike the ones that would be used in a typical high-quality dub. We aim to address this limitation in future work through specially constructed training data to reduce the reliance on accurate isochronous MT or a human-in-the-loop. Furthermore, we aim to enhance this approach by incorporating more fine-grained lip-dynamic information from the source audio, for instance by extracting phonetic content and mapping it to corresponding articulatory configurations.

\vspace{0.5cm}
\noindent\textbf{Disclosure of Generative AI Tool Use\\}
Generative AI tools were used to assist with minor grammatical corrections and stylistic improvements. These were not used to generate scientific content, results, analysis, or interpretations.

\bibliographystyle{IEEEtran}
\bibliography{bibliography} 

@article{mann1947test,
  title        = {{On a test of whether one of two random variables is stochastically larger than the other}},
  author       = {Mann, Henry B and Whitney, Donald R},
  year         = {1947},
  journal      = {The Annals of Mathematical Statistics},
  publisher    = {JSTOR},
  pages        = {50--60}
}

@article{jia2018transfer,
  title        = {{Transfer learning from speaker verification to multispeaker text-to-speech synthesis}},
  author       = {Jia, Ye and Zhang, Yu and Weiss, Ron and Wang, Quan and Shen, Jonathan and Ren, Fei and Nguyen, Patrick and Pang, Ruoming and Lopez Moreno, Ignacio and Wu, Yonghui and others},
  year         = {2018},
  journal      = {Advances in Neural Information Processing Systems},
  volume       = {31}
}

@inproceedings{wang2018style,
  title        = {{Style Tokens: Unsupervised style modeling, control and transfer in end-to-end speech synthesis}},
  author       = {Wang, Yuxuan and Stanton, Daisy and Zhang, Yu and Ryan, RJ-Skerry and Battenberg, Eric and Shor, Joel and Xiao, Ying and Jia, Ye and Ren, Fei and Saurous, Rif A},
  year         = {2018},
  booktitle    = {International Conference on Machine Learning (ICML)},
  pages        = {5180--5189},
  organization = {PMLR}
}

@inproceedings{zen2019libritts,
  title        = {{LibriTTS: A Corpus Derived from LibriSpeech for Text-to-Speech}},
  author       = {Zen, Heiga and Dang, Viet and Clark, Rob and Zhang, Yu and Weiss, Ron J and Jia, Ye and Chen, Zhifeng and Wu, Yonghui},
  year         = {2019},
  booktitle    = {{Interspeech}},
  organization = {ISCA}
}

@inproceedings{lakew-etal-2019-controlling,
  title        = {{Controlling the Output Length of Neural Machine Translation}},
  author       = {Lakew, S. M.  and Di Gangi, M. A.  and Federico, M.},
  year         = {2019},
  booktitle    = {Proceedings of the 16th International Conference on Spoken Language Translation},
  publisher    = {Association for Computational Linguistics}
}

@inproceedings{ZhangWZWCSJRR19,
  title        = {{Learning to Speak Fluently in a Foreign Language: Multilingual Speech Synthesis and Cross-Language Voice Cloning.}},
  author       = {Zhang, Yu and Weiss, Ron J. and Zen, Heiga and Wu, Yonghui and Chen, Zhifeng and Skerry-Ryan, R. J. and Jia, Ye and Rosenberg, Andrew and Ramabhadran, Bhuvana},
  year         = {2019},
  booktitle    = {Interspeech},
  publisher    = {ISCA},
  pages        = {2080--2084}
}

@inproceedings{bredin2020pyannote,
  title        = {{Pyannote. audio: neural building blocks for speaker diarization}},
  author       = {Bredin, Herv{\'e} and Yin, Ruiqing and Coria, Juan Manuel and Gelly, Gregory and Korshunov, Pavel and Lavechin, Marvin and Fustes, Diego and Titeux, Hadrien and Bouaziz, Wassim and Gill, Marie-Philippe},
  year         = {2020},
  booktitle    = {IEEE International Conference on Acoustics, Speech and Signal Processing (ICASSP)},
  pages        = {7124--7128},
  organization = {IEEE}
}

@inproceedings{liudelightfultts,
  title        = {{DelightfulTTS: The Microsoft Speech Synthesis System for Blizzard Challenge 2021}},
  author       = {Liu, Yanqing and Xu, Zhihang and Wang, Gang and Chen, Kuan and Li, Bohan and Tan, Xu and Li, Jinzhu and He, Lei and Zhao, Sheng},
  year         = {2021},
  booktitle    = {Proc. Blizzard 2021},
}

@inproceedings{ho2021classifier,
  title        = {{Classifier-Free Diffusion Guidance}},
  author       = {Ho, Jonathan and Salimans, Tim},
  year         = {2021},
  booktitle    = {NeurIPS 2021 Workshop on Deep Generative Models and Downstream Applications}
}

@inproceedings{federico-prosodic-alignment,
  title        = {{Improvements to Prosodic Alignment for Automatic Dubbing}},
  author       = {Virkar, Yogesh and Federico, Marcello and Enyedi, Robert and Barra-Chicote, Roberto},
  year         = {2021},
  booktitle    = {IEEE International Conference on Acoustics, Speech and Signal Processing (ICASSP)},
  pages        = {7543--7574},
  organization = {IEEE}
}

@article{zeghidour2021soundstream,
  title        = {{SoundStream: An end-to-end neural audio codec}},
  author       = {Zeghidour, Neil and Luebs, Alejandro and Omran, Ahmed and Skoglund, Jan and Tagliasacchi, Marco},
  year         = {2021},
  journal      = {IEEE/ACM Transactions on Audio, Speech, and Language Processing},
  publisher    = {IEEE},
  volume       = {30},
  pages        = {495--507}
}

@article{hu2021neural,
  title        = {{Neural Dubber: Dubbing for videos according to scripts}},
  author       = {Hu, Chenxu and Tian, Qiao and Li, Tingle and Yuping, Wang and Wang, Yuxuan and Zhao, Hang},
  year         = {2021},
  journal      = {Advances in Neural Information Processing Systems},
  volume       = {34},
}

@inproceedings{salesky2021multilingual,
  title        = {{The Multilingual TEDx Corpus for Speech Recognition and Translation}},
  author       = {Salesky, Elizabeth and Wiesner, Matthew and Bremerman, Jacob and Cattoni, Roldano and Negri, Matteo and Turchi, Marco and Oard, Douglas W and Post, Matt},
  year         = {2021},
  booktitle      = {Interspeech},
  pages        = {3655--3659},
  organization = {ISCA}
}

@inproceedings{tam22_interspeech,
  title        = {{Isochrony-Aware Neural Machine Translation for Automatic Dubbing}},
  author       = {{D. Tam and S. M. Lakew and Y. Virkar and P. Mathur and M. Federico}},
  year         = {{2022}},
  booktitle    = {{Interspeech}},
  organization = {ISCA}
}

@inproceedings{wilken-matusov-2022-appteks,
  title        = {{{A}pp{T}ek{'}s Submission to the {IWSLT} 2022 Isometric Spoken Language Translation Task}},
  author       = {Wilken, Patrick  and Matusov, Evgeny},
  year         = {2022},
  booktitle    = {Proceedings of the 19th International Conference on Spoken Language Translation (IWSLT 2022)},
  publisher    = {Association for Computational Linguistics},
  pages        = {369--378},
  doi          = {10.18653/v1/2022.iwslt-1.34}
}

@inproceedings{wu2022adaspeech,
  title        = {{AdaSpeech 4: Adaptive Text to Speech in Zero-Shot Scenarios}},
  author       = {Wu, Yihan and Tan, Xu and Li, Bohan and He, Lei and Zhao, Sheng and Song, Ruihua and Qin, Tao and Liu, Tie-Yan},
  year         = {2022},
  booktitle    = {Interspeech},
  pages        = {2568--2572},
  organization = {ISCA}
}

@inproceedings{koch2022poetictts,
  title        = {{PoeticTTS - Controllable Poetry Reading for Literary Studies}},
  author       = {Koch, Julia and Lux, Florian and Schauffler, Nadja and Bernhart, Toni and Dieterle, Felix and Kuhn, Jonas and Richter, Sandra and Viehhauser, Gabriel and Thang Vu, Ngoc},
  year         = {2022},
  booktitle    = {Interspeech},
  pages        = {1223--1227},
  organization = {ISCA}
}

@inproceedings{lux2023exact,
  title        = {{Exact prosody cloning in zero-shot multispeaker text-to-speech}},
  author       = {Lux, Florian and Koch, Julia and Vu, Ngoc Thang},
  year         = {2022},
  booktitle    = {IEEE Spoken Language Technology Workshop (SLT)},
  pages        = {962--969},
  organization = {IEEE}
}

@inproceedings{lipmanflow,
  title        = {{Flow Matching for Generative Modeling}},
  author       = {Lipman, Yaron and Chen, Ricky TQ and Ben-Hamu, Heli and Nickel, Maximilian and Le, Matthew},
  year         = {2023},
  booktitle    = {The Eleventh International Conference on Learning Representations (ICLR)}
}

@inproceedings{peebles2023scalable,
  title        = {{Scalable diffusion models with transformers}},
  author       = {Peebles, William and Xie, Saining},
  year         = {2023},
  booktitle    = {Proceedings of the IEEE/CVF International Conference on Computer Vision},
  pages        = {4195--4205}
}

@inproceedings{Koizumi2023,
  title        = {{LibriTTS-R: A Restored Multi-Speaker Text-to-Speech Corpus}},
  author       = {Koizumi, Yuma and Zen, Heiga and Karita, Shigeki and Ding, Yifan and Yatabe, Kohei and Morioka, Nobuyuki and Bacchiani, Michiel and Zhang, Yu and Han, Wei and Bapna, Ankur},
  year         = {2023},
  booktitle    = {{Interspeech}},
  organization = {ISCA}
}

@article{brannon2023dubbing,
  title        = {{Dubbing in Practice: A Large Scale Study of Human Localization With Insights for Automatic Dubbing}},
  author       = {Brannon, William and Virkar, Yogesh and Thompson, Brian},
  year         = {2023},
  journal      = {Transactions of the Association for Computational Linguistics},
  publisher    = {MIT Press},
  volume       = {11},
  pages        = {419--435}
}

@inproceedings{meyer2023prosody,
  title        = {{Prosody is not identity: A speaker anonymization approach using prosody cloning}},
  author       = {Meyer, Sarina and Lux, Florian and Koch, Julia and Denisov, Pavel and Tilli, Pascal and Vu, Ngoc Thang},
  year         = {2023},
  booktitle    = {International Conference on Acoustics, Speech and Signal Processing (ICASSP)},
  pages        = {1--5},
  organization = {IEEE}
}

@inproceedings{sahipjohn2024dubwise,
  title        = {{DubWise: Video-Guided Speech Duration Control in Multimodal LLM-based Text-to-Speech for Dubbing}},
  author       = {Sahipjohn, Neha and Gudmalwar, Ashishkumar and Shah, Nirmesh and Wasnik, Pankaj and Shah, Rajiv Ratn},
  year         = {2024},
  booktitle    = {Interspeech},
  pages        = {2960--2964},
  organization = {ISCA}
}

@inproceedings{eskimez2024e2,
  title        = {{E2 TTS: Embarrassingly easy fully non-autoregressive zero-shot TTS}},
  author       = {Eskimez, Sefik Emre and Wang, Xiaofei and Thakker, Manthan and Li, Canrun and Tsai, Chung-Hsien and Xiao, Zhen and Yang, Hemin and Zhu, Zirun and Tang, Min and Tan, Xu and others},
  year         = {2024},
  booktitle    = {IEEE Spoken Language Technology workshop (SLT)},
  organization = {IEEE}
}

@inproceedings{minixhofer2024ttsds,
  title        = {{TTSDS-Text-to-Speech Distribution Score}},
  author       = {Minixhofer, Christoph and Klejch, Ond{\v{r}}ej and Bell, Peter},
  year         = {2024},
  booktitle    = {IEEE Spoken Language Technology workshop (SLT)}
}

@inproceedings{minixhofer2026ttsds,
  title        = {{TTSDS}2: Resources and Benchmark for Evaluating Human-Quality Text to Speech Systems},
  author       = {Christoph Minixhofer and Ondrej Klejch and Peter Bell},
  booktitle    = {The Fourteenth International Conference on Learning Representations (ICLR)},
  year         = {2026},
}

@inproceedings{ju2024naturalspeech,
  title        = {{NaturalSpeech 3: Zero-Shot Speech Synthesis with Factorized Codec and Diffusion Models}},
  author       = {Ju, Zeqian and Wang, Yuancheng and Shen, Kai and Tan, Xu and Xin, Detai and Yang, Dongchao and Liu, Eric and Leng, Yichong and Song, Kaitao and Tang, Siliang and others},
  year         = {2024},
  booktitle    = {International Conference on Machine Learning (ICML)},
  organization = {PMLR}
}

@inproceedings{chen2024eres2netv2,
  title        = {{ERes2NetV2: Boosting Short-Duration Speaker Verification Performance with Computational Efficiency}},
  author       = {Chen, Yafeng and Zheng, Siqi and Wang, Hui and Cheng, Luyao and and others},
  year         = {2024},
  booktitle    = {Interspeech},
  organization = {ISCA}
}

@inproceedings{mehta2024matcha,
  title        = {{Matcha-TTS: A fast TTS architecture with conditional flow matching}},
  author       = {Mehta, Shivam and Tu, Ruibo and Beskow, Jonas and Sz{\'e}kely, {\'E}va and Henter, Gustav Eje},
  year         = {2024},
  booktitle    = {IEEE International Conference on Acoustics, Speech and Signal Processing (ICASSP)},
  organization = {IEEE}
}

@inproceedings{chen2025f5,
  title        = {{F5-TTS: A fairytaler that fakes fluent and faithful speech with flow matching}},
  author       = {Chen, Yushen and Niu, Zhikang and Ma, Ziyang and Deng, Keqi and Wang, Chunhui and JianZhao, JianZhao and Yu, Kai and Chen, Xie},
  year         = {2025},
  booktitle    = {Proceedings of the 63rd Annual Meeting of the Association for Computational Linguistics (ACL)},
  pages        = {6255--6271}
}

@article{zhu2025zipvoice,
  title        = {{ZipVoice: Fast and high-quality zero-shot text-to-speech with flow matching}},
  author       = {Zhu, Han and Kang, Wei and Yao, Zengwei and Guo, Liyong and Kuang, Fangjun and Li, Zhaoqing and Zhuang, Weiji and Lin, Long and Povey, Daniel},
  year         = {2025},
  journal      = {arXiv:2506.13053}
}

@article{yang2025simplespeech,
  title        = {{SimpleSpeech 2: Towards simple and efficient text-to-speech with flow-based scalar latent transformer diffusion models}},
  author       = {Yang, Dongchao and Huang, Rongjie and Wang, Yuanyuan and Guo, Haohan and Chong, Dading and Liu, Songxiang and Wu, Xixin and Meng, Helen},
  year         = {2025},
  journal      = {IEEE Transactions on Audio, Speech and Language Processing},
  publisher    = {IEEE}
}

@article{wang2025syncvoice,
  title        = {{SyncVoice: Towards Video Dubbing with Vision-Augmented Pretrained TTS Model}},
  author       = {Wang, Kaidi and He, Yi and Guan, Wenhao and Wu, Weijie and Ding, Hongwu and Zhang, Xiong and Wu, Di and Meng, Meng and Luan, Jian and Li, Lin and others},
  year         = {2025},
  journal      = {arXiv:2512.05126}
}

@article{wang2026asvspoof,
  title        = {{Asvspoof 5: Design, collection and validation of resources for spoofing, deepfake, and adversarial attack detection using crowdsourced speech}},
  author       = {Wang, Xin and Delgado, H{\'e}ctor and Tak, Hemlata and Jung, Jee-weon and Shim, Hye-jin and Todisco, Massimiliano and Kukanov, Ivan and Liu, Xuechen and Sahidullah, Md and Kinnunen, Tomi and others},
  year         = {2026},
  journal      = {Computer Speech \& Language},
  publisher    = {Elsevier},
  volume       = {95},
  pages        = {101825}
}

\end{document}